# Thermal Analysis of 3D Associative Processor

Leonid Yavits, Amir Morad, Ran Ginosar

Department of Electrical Engineering, Technion - Israel Institute of Technology, Haifa, Israel

**Abstract**— Thermal density and hot spots limit three-dimensional (3D) implementation of massively-parallel SIMD processors and prohibit stacking DRAM dies above them. This study proposes replacing SIMD by an Associative Processor (AP). AP exhibits close to uniform thermal distribution with reduced hot spots. Additionally, AP may outperform SIMD processor when the data set size is sufficiently large, while dissipating less power. Comparative performance and thermal analysis supported by simulation confirm that AP might be preferable over SIMD for 3D implementation of large scale massively parallel processing engines combined with 3D DRAM integration.

——————————— ◆ ———————————

## 1 INTRODUCTION

Machine learning, data mining, network routing, search engines and other big data applications can be significantly sped up by massively parallel SIMD machines, such as GPUs [4]. However data transfer between processing units (PUs) and memory significantly limits the performance of SIMD architectures [3]. The scalability of SIMD architectures is therefore limited by off-chip memory bandwidth constraints. To overcome this limitation, three-dimensional (3D) integrated circuits that stack DRAM over the processor are proposed; they can bring the memory much closer to the SIMD processor [6]. Additionally, multiple layers of SIMD processors may be stacked, facilitating closely-coupled parallelism. Nevertheless, thermal issues can make this approach difficult to materialize.

When operating at high rates, arrays of computing elements in SIMD processors are highly active, resulting in significant power density and hotspots [14] and creating additional design constraints such as heat dissipation, power delivery and excessive leakage [15]. Hot spots and irregular thermal density further limit the scalability of conventional SIMD architectures. These thermal considerations can impact the performance and reliability of 3D circuits [7] and render 3D integration of DRAM and SIMD processors infeasible.

Associative Processors (AP) [13][25] may offer a viable alternative to conventional SIMD processors. The AP (Fig. 1), comprising a modified Content Addressable Memory (CAM), facilitates processing in addition to content addressable and random access. Compared to SIMD, we show that APs demonstrate a more uniform thermal density and lower peak temperatures, enabling multilayer processor stacking and 3D DRAM integration.

In this study we propose to replace a massively parallel SIMD processor by an AP, particularly in 3D implementations. The goals we set to achieve are:
- Improve thermal density, reduce peak temperature, eliminate or considerably reduce hot spots.
- Combine data storage and data processing;
- Reduce performance degradation caused by massive data transfers between SIMD PUs and memory;
- Reduce energy consumed by data transfers;
- Enable multilayer processor stacking as well as 3D DRAM integration on top of the processor layers.

Achieving these goals will help enable 3D implementations combining multilayer massively parallel processing engines and multilayer DRAM.

The rest of this paper is organized as follows. Section 2 introduces the AP. Section 3 presents modeling and comparative analysis of the AP's and the SIMD processor's performance and power. Section 4 provides comparative thermal analysis of SIMD and AP, and Section 5 offers conclusions.

## 2 THE AP

In this section we present the architecture of the AP and explain the principles of associative computing.

### 2.1 AP Architecture

The AP design is based on that of a CAM. The CAM allows comparing all data words to a key, tagging the matching words, and possibly reading some or all of the tagged words one by one. In addition, normal memory read and write operations of a single word at a time can also take place.

The AP enhances the CAM by allowing parallel writing into selected bits of all tagged words. The architecture of AP is presented in Fig. 1. The Associative Processing Array comprises bit cells (further described below) organized in bit-columns and word-rows. Typically, a word (row) makes a Processing Unit (PU) (although parts of a row, or alternatively multiple rows, may also be configured as a PU). Several special registers are appended to the associative processing array. The KEY register contains a key data word to be written or compared against. The MASK register defines the active fields for write and read operations, enabling bit selectivity. The TAG register marks the rows that are matched by the compare opera-tion and may be affected by parallel write. The AP may require a microcontroller and an instruction cache. An optional Interconnect allows PUs of the AP to communicate in parallel. Since associative processing operation is mainly bitwise, the Interconnect can be a relatively simple circuit-switched network. The Interconnect is further discussed in Section 2.2.

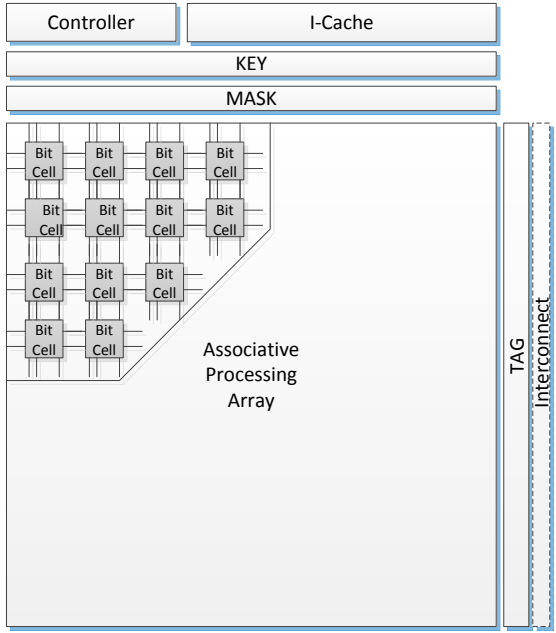

Fig. 1. AP architecture

A static memory based associative bit cell is shown in Fig. 2. Its two main components are the 6-Transistors (6T) SRAM bit cell and the 4T N-type XOR. Two additional transistors (gated by the Mask wire) are used to mask the write operation at the bit (column) level. Alternative designs have also been proposed, to reduce power dissipation [9], to save area [16] or to exploit non-transistor technology [8].

To compare the key to the data stored in the associative memory (the entire row, a number of bits or a single bit cell), the Match line is precharged and the inverted key is set on Bit and Bit-not lines. In the columns that should be ignored during comparison, Bit and Bit-not lines are set to 0. If all unmasked bits in a row match the key, the Match line remains high and a 1 is written into the corresponding TAG bit. If at least one bit is not matched, the Match line discharges and 0 is written into the TAG bit.

In AP, compare is typically followed by a parallel write into the unmasked bits of all tagged words. To write data (from the KEY register) into the associative memory, each tag bit (set earlier by the compare) is connected to the corresponding Word line. If a row has matched during the compare, the KEY data is written into it in accordance with the MASK pattern. Otherwise (in the case of mismatch), the write does not affect the row. Typically, 12.5-25% of the rows are written during a write in arithmetic operations as further shown in Section 2.2. To read data from memory, the Bit and Bit-not lines are precharged and the Word line is asserted. Parallel write and sequential read operations are enabled only for the columns whose mask bits are set in the MASK register.

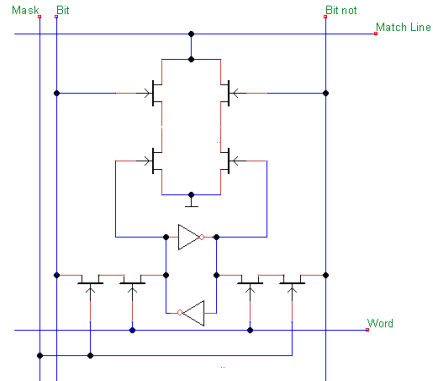

Fig. 2. NOR-type Associative Bit Cell

### 2.2 Associative Computing

AP is a general purpose computational device that can implement a wide range of arithmetic, logic and processing tasks in addition to classical CAM operations such as associative search, sorting and ordering, and in addition to standard memory operations (word and block read and write).

Arithmetic operations in the AP can be performed in parallel on all PUs in a word-parallel, bit-serial manner. For instance, vector addition may be performed as follows [1]. Two $m$ bit columns hold vectors A and B (Fig. 3). Their sum A+B is written over B. A one-bit column C holds the carry bit. The addition is carried out in $m$ single-bit addition parallel steps (1):

$$c[*] \mid s[*]_i = a[*]_i + b[*]_i + c[*] \quad (1)$$
$$\forall\, i = 0, \ldots, m-1$$

where $i$ is the bit index and '$*$' is the word index in the vector. The single-bit addition (TABLE 1) is carried out in a series of passes, where in each pass, one entry of the truth table (a three bit input pattern) is matched against the contents of the $a[*]_i, b[*]_i, c[*]$ bit columns and the matching rows (PUs) are tagged; then the logic result (two-bit output of the truth table in TABLE 1) is written into $b_i$ and $c$ bits of all tagged rows. During this operation, all but three input bit columns and two output bit columns of the associative array are masked out in each pass, so that 2.5 bit columns are active on average. Some input combinations do not change the output and therefore can be skipped ("No action" in the table). Since the operation overwrites one of the inputs, computation must be carried out according to the order indicated in the table.

Overall, four passes of one compare and one write operations are required to complete the single-bit addition. Therefore, fixed point $m$ bit addition takes $8m \in O(m)$ cycles. Subtraction and comparison operations are performed similarly and also require $O(m)$ cycles. Notice the stark contrast with SIMD architectures of low PU count that require $O(N)$ cycles to add $N$ data elements (without taking into account the load / store time).

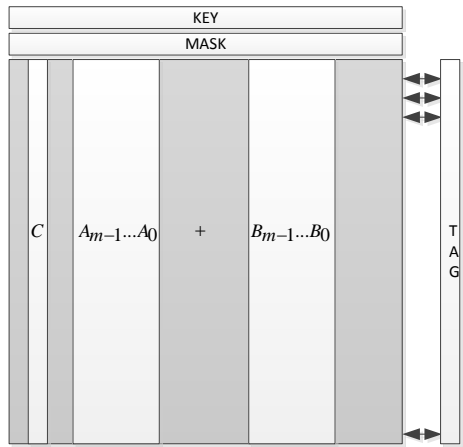

Fig. 3. Addition Example

TABLE 1
IMPLEMENTING FULL ADDER IN AP

| Entry | Input C | Input B | Input A | Output C | Output B | Comments |
|---|---|---|---|---|---|---|
| 0 | 0 | 0 | 0 | 0 | 0 | No action |
| 1 | 0 | 0 | 1 | 0 | 1 | 2nd pass |
| 2 | 0 | 1 | 0 | 0 | 1 | No action |
| 3 | 0 | 1 | 1 | 1 | 0 | 1st pass |
| 4 | 1 | 0 | 0 | 0 | 1 | 3rd pass |
| 5 | 1 | 0 | 1 | 1 | 0 | No action |
| 6 | 1 | 1 | 0 | 1 | 0 | 4th pass |
| 7 | 1 | 1 | 1 | 1 | 1 | No action |

*Pass = COMPARE cycle followed by WRITE cycle*

Fixed precision multiplication and division in AP are implemented by long multiplication and division respectively, consisting of a series of add-shift and subtract-shift operations, executed bit-serially but word-parallel. The addition or subtraction are done as described above (multiplication is usually done "MSB first"), while shift is implemented by activating different bit columns and therefore requires no cycles. Thus, fixed point $m-bit \times m-bit$ vector multiplication requires $O(m^2)$ cycles [1] regardless of the length of the vectors.

Floating point arithmetic for APs is somewhat more complex to implement. Different exponents require shifting mantissas by different lengths, resulting in a sequence of bit-serial operations. Still, a direct implementation of IEEE single precision floating point vector multiplication requires only 4400 cycles, regardless of the length of the vector.

In general, any computational expression can be efficiently implemented on an AP using this look up table (LUT) approach, where all possible arguments of the function are matched with the contents of the associative memory, and the corresponding function values are written in the designated fields of the tagged memory rows. For a $m$-bit argument $x$, such $f(x)$ has $2^m$ possible values, therefore the LUT operation incurs $O(2^m)$ cycles. Obviously, all values of the $f(x)$ LUT are pre-calculated and implicitly stored in AP instructions.

Arithmetic operations are presented in this Section under the assumption that all operands are located in the PU. However, many workloads require inter-PU data communications. Depending on the workload, communication requirements may vary from no communications (for "embarrassingly parallel" tasks such as Black-Scholes option pricing) to relatively intense communications (e.g., for dense matrix multiplication and other linear algebra tasks). In some cases, support for special pre-defined communication patterns or permutations can be of advantage (e.g., for FFT). The inter-PU communication can be implemented serially, through a series of associative memory reads and writes. Alternatively, the dedicated Interconnect introduced in Section 2.1 can be employed to provide parallel communication capabilities, i.e. to allow all PUs to communicate in parallel.

## 3 AP VS. SIMD PROCESSOR

### 3.1 Performance

For the comparison of the thermal characteristics of the AP vs. the SIMD processor to be meaningful, both processors need to produce the same performance (i.e. complete the same task in the same amount of time). We employ a model that predicts performance and power based on area, and compare SIMD and AP chips of appropriate areas that deliver the same level of performance. Thermal analysis is then based on the resulting area and power figures.

Performance is evaluated through simulations [19], using the following single precision floating point workloads:
- $N$-option pairs Black-Sholes option pricing (BS)
- $N$-point Fast Fourier Transform (FFT)
- $\sqrt{N} \times \sqrt{N}$ Dense Matrix Multiplication (DMM)

where $N = 2^{20}$. These workloads are significant because they span a wide range of computation-to-bandwidth ratio, also referred to as 'arithmetic intensity' (Fig. 4).

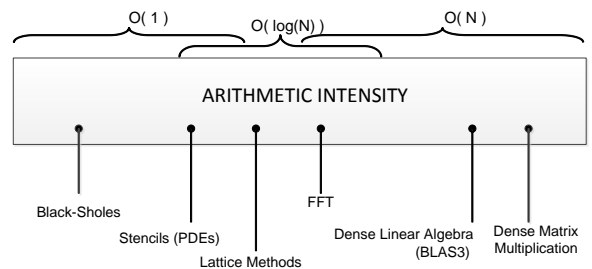

Fig. 4: Arithmetic Intensity [23]

For dense matrix multiplication and FFT, we used optimized implementations outlined in [21]. For Black-Scholes, we used a direct implementation optimized for associative processing, based on formulation in [5].

The reference SIMD architecture is presented in Fig. 5. It consists of twelve identical processors, each containing 64 parallel PUs with register files and a L1 cache (the analysis below determines that the SIMD should include 768 PUs). Additionally, the reference SIMD processor features a L2 cache shared by all twelve processors.

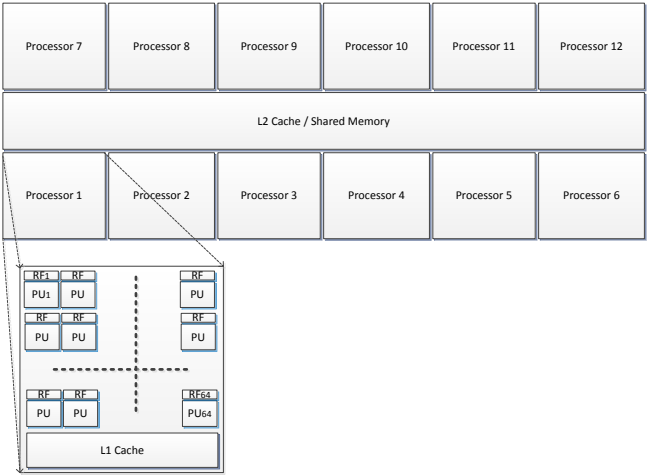

Fig. 5. Reference SIMD architecture (12 × 64 PUs)

Consider the foregoing model predicting performance as a function of area. Let the serial execution time (on a single PU of the SIMD processor) of a workload be $T_1$. The execution time $T_{SIMD}$ of the same workload on the SIMD architecture of Fig. 5 is bounded by:

$$T_{SIMD} \geq \frac{T_1}{n_{SIMD}} + T_S \quad (2)$$

where $n_{SIMD}$ is the number of PUs (shown below to be 768) and $T_S$ is the caches-to-PU synchronization time spent exclusively on the data transfer between L1 and L2 caches and the PUs of the SIMD processor. The maximum speedup (which we use as the measure of performance) of the SIMD processor over a single PU can be written as follows:

$$S_{SIMD} = \frac{T_1}{T_{SIMD}} = \frac{T_1}{\frac{T_1}{n_{SIMD}} + T_S} = \frac{1}{\frac{1}{n_{SIMD}} + I_s} \quad (3)$$

where $I_s = T_S/T_1$ is the *synchronization intensity*, or the ratio of time spent on caches-to-PU synchronization to the serial execution time. Synchronization intensity is inversely proportional to the arithmetic intensity of Fig. 4.

The area of the SIMD processor can be presented as follows:

$$A_{SIMD} = n_{SIMD}(A_{PU} + A_{RF}) + A_C \quad (4)$$

where $A_{PU}$ is the PU area, $A_{RF}$ is the register file area, and $A_C$ is the area of the L1 and L2 cache memories (of total size of at least $N$ data words). For easy comparison between PU and memory areas, we represent all area values (logic and memory) in terms of a baseline SRAM bit cell area, which we assume to be 1. In current CMOS technology, the actual figure is in about $0.1\mu m^2$. Then we can write:

$$A_{PU} = A_{PUo}m^2$$
$$A_{RF} = A_{RFo}km \quad (5)$$

where $A_{PUo}$ is the area of a single bit of the PU and $A_{RFo}$ is the area of a register bit (a flip-flop), both measured in baseline SRAM bit cell area units; $m$ is data word-length and $k$ is the depth of the register file (see TABLE 2).

We can derive $n_{SIMD}$, the number of PUs in the SIMD processor, as a function of its total area $A$ using (4) and (5) as follows:

$$n_{SIMD} = \frac{A - A_C}{A_{PU} + A_{RF}} = \frac{A - A_C}{A_{PUo}m^2 + A_{RFo}km} \quad (6)$$

For simplicity we assume that $A_C, A_{PUo}, A_{RFo}, k$ and $m$ (TABLE 2) are constants and do not change with total area $A$. We further substitute $n_{SIMD}$ in (3) by (6) and receive the speedup of the reference SIMD processor as a function of its area. This simplified model does not take into account many aspects of the SIMD processor design; its purpose is only to provide best case reference figures for the comparative analysis.

The execution time of a workload on the AP can be written as follows:

$$T_{AP} = \frac{T_1}{s_{APU}n_{AP}} \quad (7)$$

where $n_{AP}$ is the number of PUs in the AP and $s_{APU}$ is the speedup of the associative PU relative to the SIMD PU. The lower bound of $s_{APU}$ is 1/4400, which is the ratio of SIMD to AP floating point multiplication times (1 cycle *vs.* 4400 cycles, cf. Section 2.2). Since the AP combines data processing and data storage, there is no need for synchronization and therefore $T_s$ is omitted from (7). The speedup of the AP can then be written:

$$S_{AP} = s_{APU}n_{AP} \quad (8)$$

The area of the AP can be written as follows:

$$A_{AP} = n_{AP}A_{APo}km \quad (9)$$

where $k$ is the size of associative PU (the number of data words per PU) and $A_{APo}$ is the AP cell area, measured in SRAM cell area units (TABLE 2).

Similarly to (6), $n_{AP}$ (the number of AP PUs) may be derived as a function of its total area using (9) as follows:

$$n_{AP} = \frac{A}{A_{APo}km} \quad (10)$$

We can now substitute $n_{AP}$ in (8) by (10) and obtain the speedup of the AP as function of its area.

The results of speedup *vs.* area simulations for three different workloads [19] are presented in Fig. 6. These simulation results were shown in [19] to be compatible with the analytical predictions of (3) and (8). Incidentally, these findings show that for each workload there is an area budget point beyond which the AP outperforms the SIMD processor (the break-even point), whose speedup is constrained by the caches-to-PU synchronization time $T_S$. The variations in simulation behavior are the result of the differences in arithmetic intensity of each individual workload.

TABLE 2
AREA MODEL PARAMETERS

| Parameter | Description | Attributed to | Value |
|---|---|---|---|
| $A_{SRAM-cell}$ | Area of SRAM cell | Both | $0.1\mu m^2$ |
| $A_{PUo}$ | PU bit cell area | SIMD | 20 [1] |
| $A_{RFo}$ | Register bit (FF) area | SIMD | 3 [1] |
| $S_{APU}$ | AP PU speedup relative to SIMD PU | AP | 1/4400 |
| $A_{APo}$ | AP bit area | AP | 2 [1] |
| $m$ | Data word-length | Both | 32 |
| $k$ | Number of data words in temporary storage per PU | Both | 8 |

[1] Area parameters are normalized to $A_{SRAM\text{-}cell}$

For thermal analysis, consider dense matrix multiplication, the most demanding workload, where SIMD processor achieves the highest speedup. We scale the AP size to the data set size, i.e. $n_{AP} = N = 2^{20}$ and $A_{AP}$=53mm². At this size, AP reaches the speedup of 350 (marked by the black dotted line in Fig. 6). To yield the same speedup, $n_{SIMD}$ (the number of PUs in the SIMD processor) should be 768 and $A_{SIMD}$=5.3mm².

Another comparison may be made at the break-even point where the two architectures not only achieve the same performance but also have the same area, such as the red circle on the FFT curves in Fig. 6.

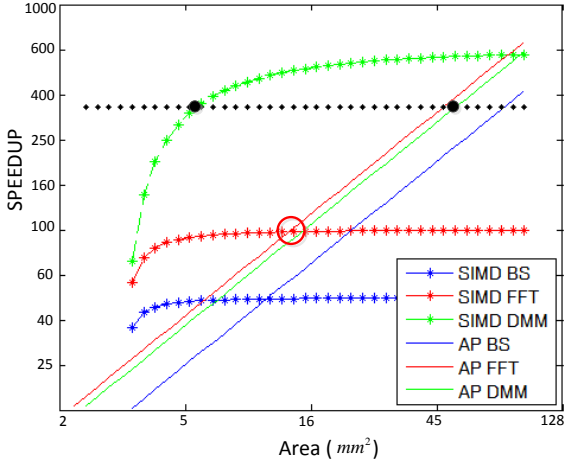

Fig. 6. Speedup vs. Area

## 3.2 Power

Thermal behavior is based on power dissipation. We employ a model that predicts power consumption of the AP and the SIMD processor based on area.

The average power of the SIMD processor (over the execution span $T_{SIMD}$) can be written as follows:

$$P_{SIMD} = \frac{E_{EXEC} + E_S + E_{LEAK}}{T_{SIMD}}$$

$$= \frac{P_{EXEC} \cdot \frac{T_1}{n_{SIMD}} + P_S \cdot T_S + P_{LEAK} \cdot (\frac{1}{n_{SIMD}} + I_s) \cdot T_1}{(\frac{1}{n_{SIMD}} + I_s) \cdot T_1} \quad (11)$$

$$= \frac{\frac{P_{EXEC}}{n_{SIMD}} + I_s \cdot P_S}{(\frac{1}{n_{SIMD}} + I_s)} + P_{LEAK}$$

where $E_{EXEC}$ and $P_{EXEC}$ are the energy and the average power consumption during execution; $E_S$ and $P_S$ are the energy and the average power consumed during caches-to-PU synchronization; $E_{LEAK}$ and $P_{LEAK}$ are the leakage energy and power; $I_s$ is the synchronization intensity as defined above.

Just as in the case of area comparison, power values are normalized to the power consumption of a baseline SRAM memory cell during a write operation, which we assume to be 1. In current high-speed CMOS technology, the actual figure is in the range of ~0.5 $\mu W$. Then we can further write the SIMD power consumption as follows:

$$P_{EXEC} = n_{SIMD}(P_{PUo}m^2 + P_{RFo}km)$$
$$P_S = P_{So}m \quad (12)$$

where $P_{PUo}$ and $P_{RFo}$ are the average per-bit power consumptions of the PU and RF respectively during execution (computation) (TABLE 3). $P_{So}$ is the power consumed performing the caches-to-PU synchronization of a single data bit. We assume the amount of data that needs to be synchronized with the caches is limited to a single data word per PU (as typical in applications such as dense matrix multiplication and FFT). $P_{EXEC}$ and $P_S$ are normalized to a single bit write into SRAM.

Leakage power can be expressed as follows:

$$P_{LEAK} = \beta A V^\alpha = \gamma A \quad (13)$$

where $A$ is the area, $V$ is the supply voltage, and α and β are constants, and γ is the leakage area coefficient that depends on silicon process and operating conditions. Therefore the total SIMD processor power can be written as follows:

$$P_{SIMD} = \frac{P_{PUo}m^2 + P_{RFo}km + I_s P_{So}m}{\frac{1}{n_{SIMD}} + I_s} + \\ + \gamma \cdot n_{SIMD} \cdot (A_{PUo}m^2 + A_{RFo}km) \quad (14)$$

The average power of the AP can be written as follows:

$$P_{AP} = \frac{E_{EXEC} + E_{LEAK}}{T_{AP}} = P_{EXEC} + P_{LEAK} \\ = [\frac{P_W + P_C}{2} + \gamma A_{APo}km]n_{AP} \quad (15)$$

where $E_{EXEC}$ and $P_{EXEC}$ are the AP execution energy and power consumption; $E_{LEAK}$ and $P_{LEAK}$ are the AP leakage energy and power; $P_W$ is the power consumption during the associative memory write and $P_C$ is the power consumption during compare (typically, AP compute time divides equally between compare and write operations). In order to further detail the dynamic power of the AP, recall the implementation of single-bit addition (on which other arithmetic operations are based) described in Section 2.2. In each pass of the single-bit addition, a three bit input combination $a[*]_i, b[*]_i, c[*]$ is compared in parallel in all PUs and afterwards a two bit result $b[*]_i, c[*]$ is written into the tagged PUs; that sequence is repeated $m$ times for $m$-bit words. Since there are eight independent logic combinations (TABLE 1), each PU has 1/8 probability of match and 7/8 of mismatch (in which case the Match line discharges). Similarly, each PU has 1/8 probability of write and 7/8 probability of a *miswrite* (when Bit and Bit-not lines are charged without Word line being asserted). Since we define the power consumption of a single SRAM cell during write operation as 1, (15) can be rewritten as:

$$P_{EXEC} = \frac{2 \cdot (1/8 \cdot 1 + 7/8 \cdot p_{mw}) + 3 \cdot (1/8 \cdot p_m + 7/8 \cdot p_{mm})}{2} n_{AP} \quad (16)$$

for 2-bit write and 3-bit compare operations, where $p_{mw}$ is the normalized per-bit power consumption of a miswrite, $p_{mm}$ is the normalized per-bit power consumption of a mismatch, and $p_m$ is the normalized per-bit power consumption of a match (TABLE 3).

Using (13), we can write the total power dissipation of the AP as follows:

$$P_{AP} = n_{AP} \cdot [\frac{1}{8} + \frac{7}{8}p_{mw} + \frac{3}{16}p_m + \frac{21}{16}p_{mm} + \gamma A_{AP_o} km] \quad (17)$$

This model is fairly basic and does not account for certain statistics that work in favor of the AP. For example, a certain percentage of associative memory cells that are written a new value in fact do not change (consuming considerably less power); similarly, a certain percentage of asserted bit lines do not recharge (or discharge) since the same value is asserted. Our goal is to create a simple power model that reflects the worst case power consumption of the AP.

Power consumption *vs.* area for SIMD and AP for the three workloads is presented in Fig. 7. The two area points selected for dense matrix multiplication in Fig. 6 are marked here by black dots for reference; notice that for the same performance, SIMD consumes more than twice the power of AP. Since the SIMD area is one tenth of the AP, the power density is about twenty five times higher. In the following section we show this effect as thermal disparity.

Notice also the two red circles in Fig. 7. They represent the two different power levels that are required for the FFT case where the two architectures achieve the same performance and also occupy the same area (at the break-even point); since the SIMD processor incurs higher power density (higher power over the same area), clearly its thermal level is also going to be higher than the AP.

TABLE 3
POWER MODEL PARAMETERS

| Parameter | Description | Attributed to | Value |
|---|---|---|---|
| $P_{SRAM-cell}$ | Power of SRAM cell during write operation | Both | 0.5 µW [1] |
| $P_{PUo}$ | per-bit power consumption of PU | SIMD | 40 [2] |
| $P_{RFo}$ | per-bit power consumption of RF | SIMD | 5 [2] |
| $P_{So}$ | per-bit power consumption during synchronization | SIMD | 200 [2] |
| $p_{mw}$ | per-bit power consumption during a miswrite | AP | 0.1 [1] [2] |
| $p_m$ | per-bit power consumption during a match | AP | 0.1 [1] [2] |
| $p_{mm}$ | per-bit power consumption during a mismatch | AP | .75 [1] [2] |
| $\gamma$ | leakage power coefficient | Both | $5 \cdot 10^{-2}$ W/mm² |

*COMPARE and WRITE are executed within the AP array; there is no data transfer outside the AP memory*
*(1) SRAM power figures are based on [2]*
*(2) Power parameters are normalized to $P_{SRAM-cell}$*

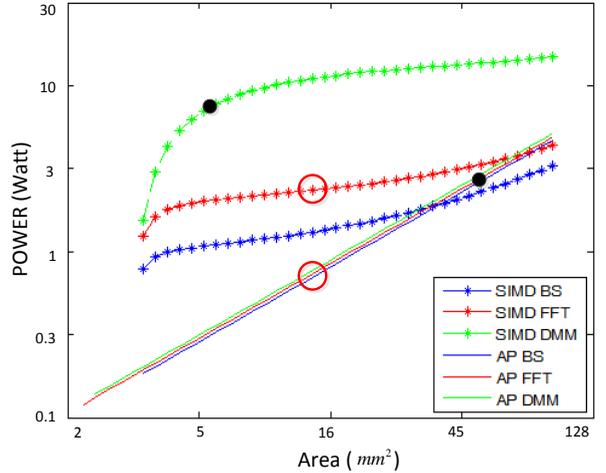

Fig. 7. Power vs. Area

## 4 3D THERMAL ANALYSIS

The purpose of thermal analysis is to compare the thermal distribution and the peak temperatures (hot spots) of the 3D AP *vs.* the 3D SIMD processor. We perform thermal analysis using the HotSpot simulator, a tool for architectural thermal modeling [14][18][17][24].

### 4.1 3D AP Thermal Analysis

The floorplan of the AP (excluding a controller and an I-cache) used for HotSpot simulation is shown in Fig. 8. The 7.3×7.3mm AP is divided into 64 identical banks (Fig. 8 (a)). Each AP bank is further divided into 64 identical blocks (Fig. 8 (b)), and each block features a 256×256 associative processing array (256 PUs, 256-bit each), 256-bit TAG register to the right of the associative processing array, and 256-bit KEY and MASK registers at the top (Fig. 8 (c)). The total number of PUs in the AP is $2^{20}$, as in Section 3.1. Fig. 9 presents the 3D thermal model of the 3D AP. The 3D stack comprises four silicon layers (each containing the AP of Fig. 8(a)) above a Thermal Interface Material (TIM) layer. Actual structures may also contain Heat Spreader (HSP) layer and a heat-sink [6].

For power trace (as needed for HotSpot simulation), we use the results of power modeling from Section 3.2.

The results of HotSpot simulation are presented in Fig. 10. Fig. 10 (c) shows the thermal map of the individual block$_{4,4}$ located near the center of bank$_{4,4}$ of the AP. Its maximum temperature reaches 55°C. The hottest part of the AP block is the KEY and the MASK register area, where 2% of flip-flops are switching every cycle. Another noticeable region is in the middle of the associative processing array. Fig. 10 (b) presents the thermal map of bank$_{4,4}$ located around the center of the AP. Its maximum temperature reaches 55°C. Fig. 10 (a) presents the thermal map of the AP placed at the top silicon layer. The peak temperature of this layer is 55°C. The hottest region of the AP is located at its center. This is quite expected since during arithmetic operations, the active (switching) elements are uniformly distributed across the AP and therefore the resulting temperature is distributed normally across the plane. Note that the temperature span in the AP is only about 3°C. Hence we can assert that thermal distribution of the AP is very close to uniform.

### 4.2 3D SIMD Processor Thermal Analysis

A reference massively parallel SIMD architecture used in thermal analysis is presented in Fig. 5 and discussed in Section 3.1. The floorplan of the reference SIMD processor is depicted in Fig. 11. Its area is 2.3×2.3mm.

For power trace (as needed for HotSpot simulation), we use the results of the power modeling presented in Section 3.2. We target the same 3D thermal model depicted in Fig. 9: four layers of SIMD processors of Fig. 11 are stacked one on top another, followed by TIM and HSP layers and a heatsink.

The results of HotSpot simulation are shown in Fig. 12, which presents the thermal map of the SIMD processor placed in the upper silicon layer. The temperature ranges from 98°C to 128°C. The hottest part of the reference SIMD processor is the PU array, and the coolest region lies around the center of the L2 cache.

### 4.3 Comparison

Fig. 13 presents the thermal distribution of the four silicon layers of the AP and the SIMD processor along the T-Cut sections (cf. Fig. 8 and Fig. 11) for the same-performance dense matrix multiplication workload. Similar results are obtained when considering the same-performance, same-area case of FFT (the circles in Fig. 6 and Fig. 7).

The peak temperature of the SIMD processor located in the upper silicon layer is above the maximal operating temperature of most commercially available DRAMs (85°C-95°C) [10][11][12], which prohibits 3D DRAM integration above the SIMD processor. Comparison of HotSpot simulation results suggests that the AP is better positioned for multilayer 3D stacking and integration with 3D DRAM than a massively parallel SIMD processor.

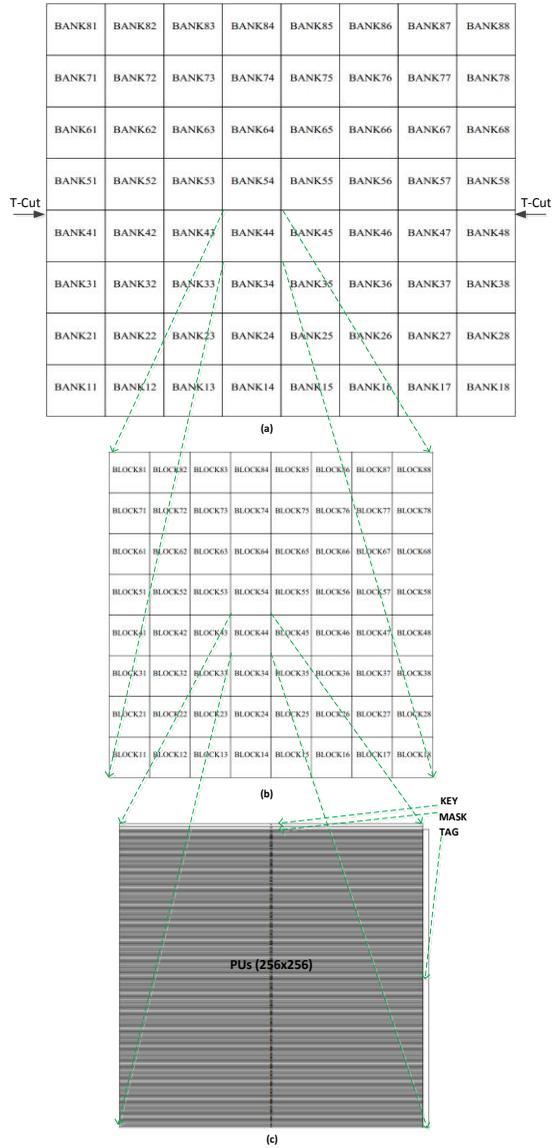

Fig. 8. AP floorplan for HotSpot thermal modeling

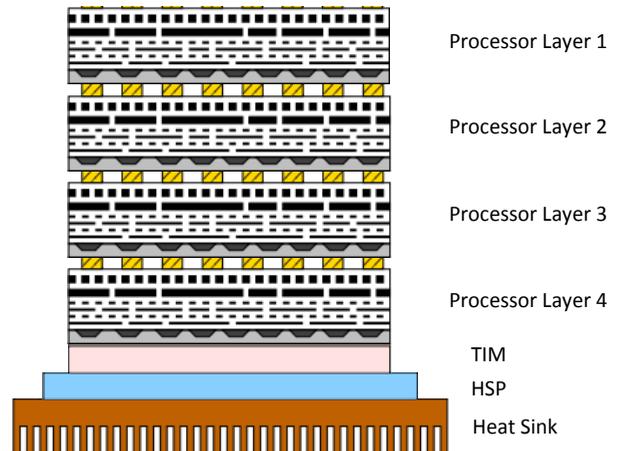

Fig. 9. 3D thermal model (based on [6])

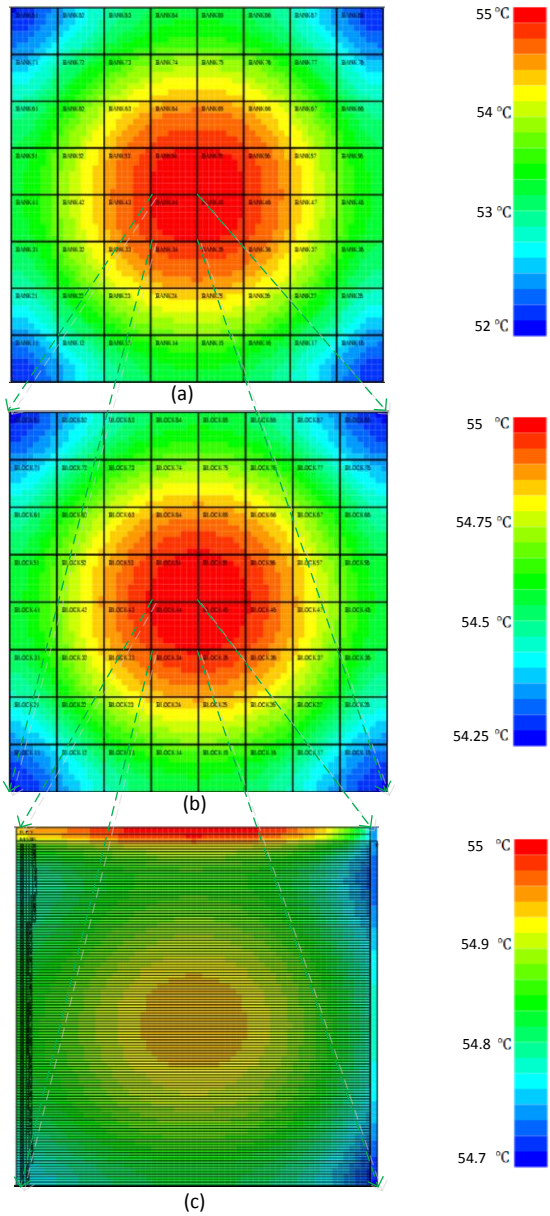

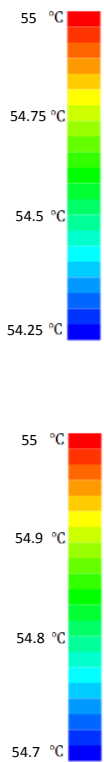

Fig. 10. A 52-55°C thermal map of the layer 1 AP

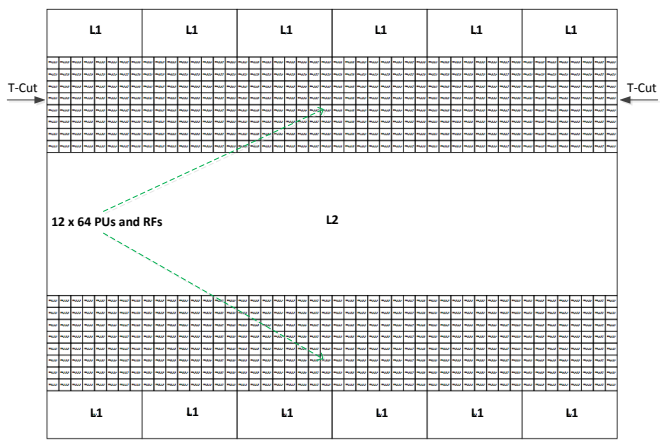

Fig. 11. Reference SIMD floorplan for HotSpot thermal modeling

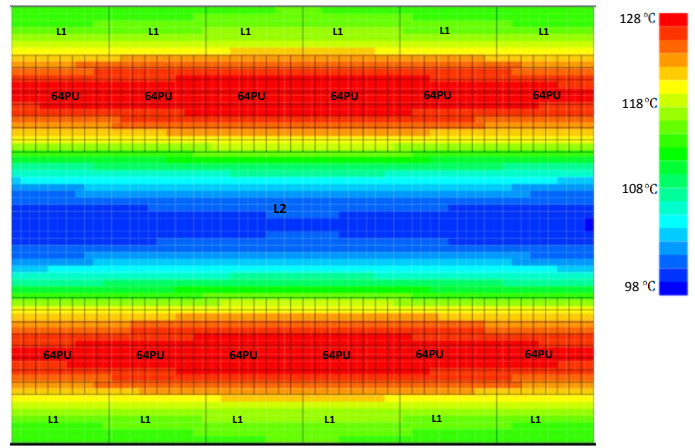

Fig. 12. A 98-128°C thermal map of the layer 1 SIMD processor

Fig. 13. AP vs. SIMD temperatures in four silicon layers along the T-Cuts (see Fig. 8 and Fig. 11)

## 5 CONCLUSIONS

The associative processor (AP) is essentially a large associative memory with massively-parallel processing capabilities. The AP exhibits thermal distribution that is close to uniform. This paper investigates the merit of 3D implementation of APs relative to SIMD processors, and shows that in addition to eliminating hot spots, AP temperature levels are significantly lower than those of SIMD processors yielding the same performance. The HotSpot thermal simulator was used on data obtained from simulations of several workloads.

Thanks to low and almost uniformly distributed thermal levels, we show that 3D AP implementations allow stacking multiple DRAM dies above multiple processor layers. In contrast, integrating 3D DRAM with a conventional massively parallel SIMD processor might be impractical due to hot spots.

Another advantage of AP is unification of processing and storage. AP enables true in-memory computing down to the bit-cell level. This study shows that the speedup of SIMD processors may be limited by massive data transfers between the processing units and the caches. That limitation becomes more restrictive as the data set size and the SIMD processor size grow. AP offers the potential for higher speedup thanks to its in-memory computation that reduces the negative effect of PU-to-caches data synchronization.

In summary, associative processing architectures can potentially enable high performance 3D stacking thanks to thermal advantages over massively parallel SIMD architectures such as high-end GPUs.

## ACKNOWLEDGMENT

We are grateful to Eby Friedman for motivating this research. This research was partly funded by the Intel Collaborative Research Institute for Computational Intelligence and by Hasso-Plattner-Institut.